\documentclass[twocolumn,trackchanges]{aastex701}

\begin{document}

\title{Minor Merger, Major Growth: An Overmassive, Highly Accreting Black Hole Powering a Secondary AGN In a Cosmic Noon Minor Merger}

\author[orcid=0000-0003-3726-5611]{Marko Mi\'ci\'c}
\affil{Homer L.\ Dodge Department of Physics and Astronomy,
University of Oklahoma, Norman, OK 73019, USA}
\email[show]{micic@ou.edu}

\begin{abstract}

We report the discovery of a spectroscopically confirmed $z = 1.824$ minor merger with a mass ratio of $\sim$35:1 in which the secondary (smaller) galaxy hosts a luminous AGN. The system is identified in the 3D-HST survey and exhibits clear tidal features in James Webb Space Telescope imaging, confirming an ongoing interaction. Using archival Chandra X-ray observations, we detect $121 \pm 11$ X-ray counts associated with the secondary galaxy, corresponding to a rest-frame $2$–$10$ keV luminosity of $L_X \approx 9\pm0.1 \times 10^{43}$ erg s$^{-1}$ and a photon index of $\Gamma \approx$ 2.0-2.3. Analysis of the HST/WFC3 G141 grism spectrum yields an [\ion{O}{3}] $\lambda5007$ luminosity of $2\pm0.5 \times 10^{42}$ erg s$^{-1}$. Independent bolometric luminosity estimates from X-ray and [\ion{O}{3}] emission are consistent, implying $L_{\rm bol} \sim (3$–$7) \times 10^{45}$ erg s$^{-1}$. Assuming standard black hole–galaxy scaling relations, the expected black hole mass is $\sim 2 \times 10^6,M_\odot$, which would require extreme super-Eddington accretion to explain the observed luminosity. On the other hand, assuming Eddington-limited or moderately sub-Eddington accretion implies a black hole mass more than an order of magnitude above expectations. The observed X-ray spectral slope disfavors low accretion rates, restricting the allowed parameter space to high $\lambda_{\rm Edd}$ and elevated black hole masses. We conclude that the secondary AGN must be powered by an overmassive, highly accreting black hole, providing direct observational support for theoretical predictions that minor mergers can drive rapid black hole growth in secondary, smaller companions.
\end{abstract}

\keywords{\uat{Galaxy interactions}{600} --- \uat{Active galactic nuclei}{16} --- \uat{Supermassive black holes}{1663}}


\section{Introduction} 
The existence of supermassive black holes (SMBHs) with masses $\gtrsim$ 10$^6$ Msol at $z>6$ represents a major challenge for models of black hole growth and evolution \citep{2018Natur.553..473B,2020ApJ...897L..14Y,2021ApJ...907L...1W,2024NatAs...8..126B}. The short available time after the Big Bang requires either massive initial seeds or sustained Eddington-limited accretion and rapid growth of lower-mass seeds to be the progenitors of the first SMBHs.

Galaxy interactions and mergers are widely considered to be efficient fueling mechanisms and key channels for accelerating black hole growth \citep{2011MNRAS.418.2043E,2013MNRAS.435.3627E,2014MNRAS.441.1297S,2018PASJ...70S..37G,2020A&A...637A..94G,2024ApJ...968L..21M}. During these galactic encounters, gravitational torques can remove angular momentum from cold gas, driving inflows toward the nuclear region and triggering episodes of enhanced accretion. While major mergers, interactions in which the involved galaxies have comparable masses, have traditionally been emphasized for their dramatic impact on galaxy morphology, they are relatively rare compared to minor, unequal-mass mergers (M$_1$/M$_1>$ 3) and therefore may not dominate the global black hole growth budget.

Simulations of minor mergers suggest that the primary AGN, which resides in the larger galaxy, is mostly insensitive to the interaction. However, a secondary AGN, residing in the smaller galaxy, reacts dramatically to the interaction. A secondary AGN in a typical 10:1 minor merger is expected to undergo three short bursts of intensive accretion at pericentric passages, potentially leading to a tenfold increase in black hole mass over a short period of time \citep{2011ApJ...729...85C,2012ApJ...748L...7V,2015MNRAS.447.2123C}. These results suggest that minor mergers could provide an efficient, previously underappreciated pathway for building SMBHs, particularly in low-mass galaxies.

Despite their importance, observational studies of secondary AGN in minor mergers remain scarce, with only a few such objects discovered to date \citep{2012ApJ...746L..22K,2015ApJ...806..219C,2017ApJ...836..183S,2018ApJ...862...29L,2024OJAp....7E...3M}. This is largely due to a combination of spatial resolution limits, selection biases against faint companions, and the difficulty of disentangling nuclear activity from star formation in low-mass systems. Additionally, almost all known secondary AGN in minor mergers have only limited multiwavelength observations, significantly limiting the science that can be extracted. 

An exception is Was49b, a $z=0.06328$ minor merger known for more than three decades to host an extremely luminous secondary AGN \citep{1983ApJ...272...68W,1989ApJS...70..271B,1992AJ....104..990M}. Follow-up Chandra and Swift observations established that the secondary nucleus, which is seven to fifteen times less massive than the primary, harbors an overmassive SMBH, accounting for $\sim2\%$ of the total stellar mass \citep{2017ApJ...836..183S}.

Here, we report a serendipitous discovery of a $\sim$35:1 minor merger at $z=1.824$ in which the secondary galaxy hosts a luminous AGN. Using Hubble Space Telescope spectroscopy and Chandra X-ray observations, we constrain the properties of the [\ion{O}{3}] $\lambda5007$ and high-energy emission, finding that the AGN is powered by an overmassive and probably highly accreting black hole. This system provides rare direct support for theoretical predictions and offers insight into the mechanisms that govern rapid black hole growth during cosmic noon.
\begin{figure*}[ht!]
\plotone{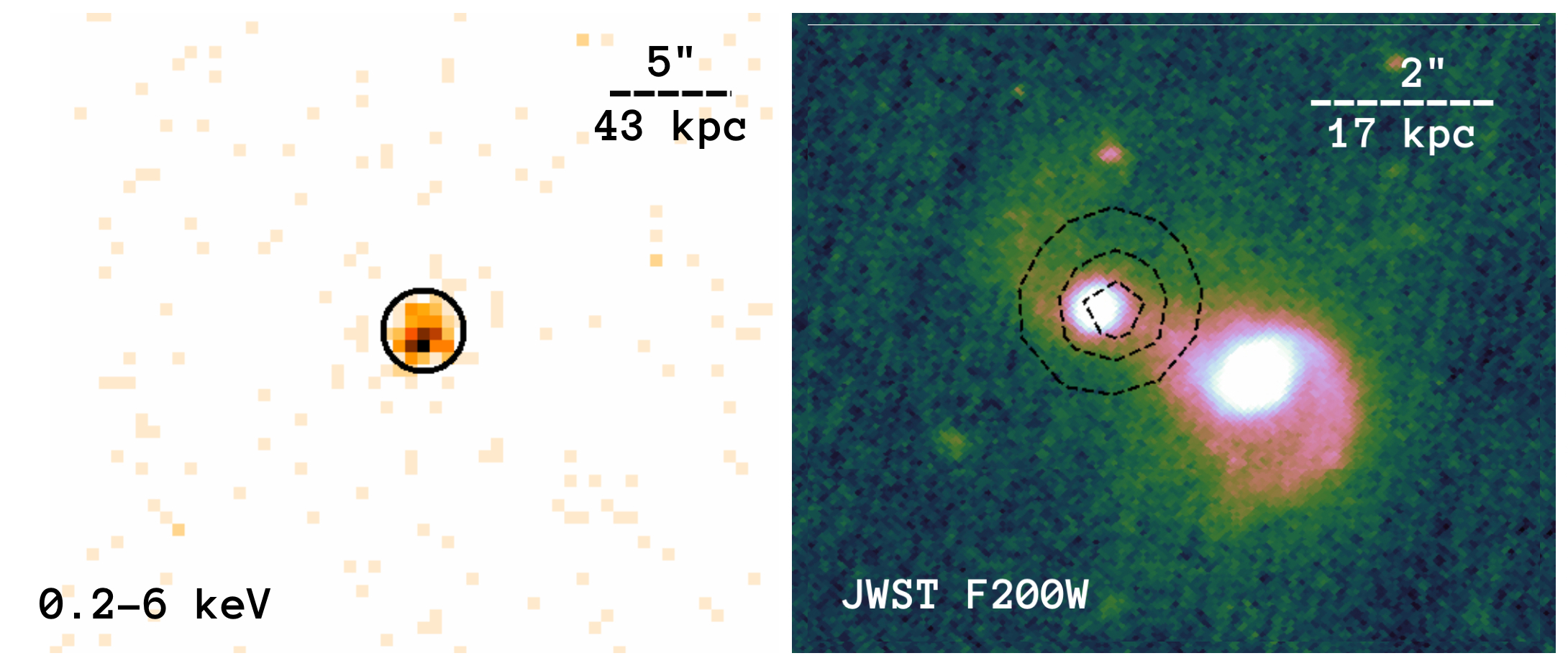}
\caption{Left: Chandra X-ray image in the 0.2-6 keV energy band. The black circle denotes a 1\farcs65-radius source-extraction region. The X-ray source is detected with 121$\pm$11 photons. Right: JWST F200W image of the 35:1 minor merger, showing the primary and secondary galaxy. The black contours indicate the location of the X-ray source, which is associated with the smaller, secondary galaxy.
\label{fig:general1}}
\end{figure*}
\begin{figure}[ht!]
\plotone{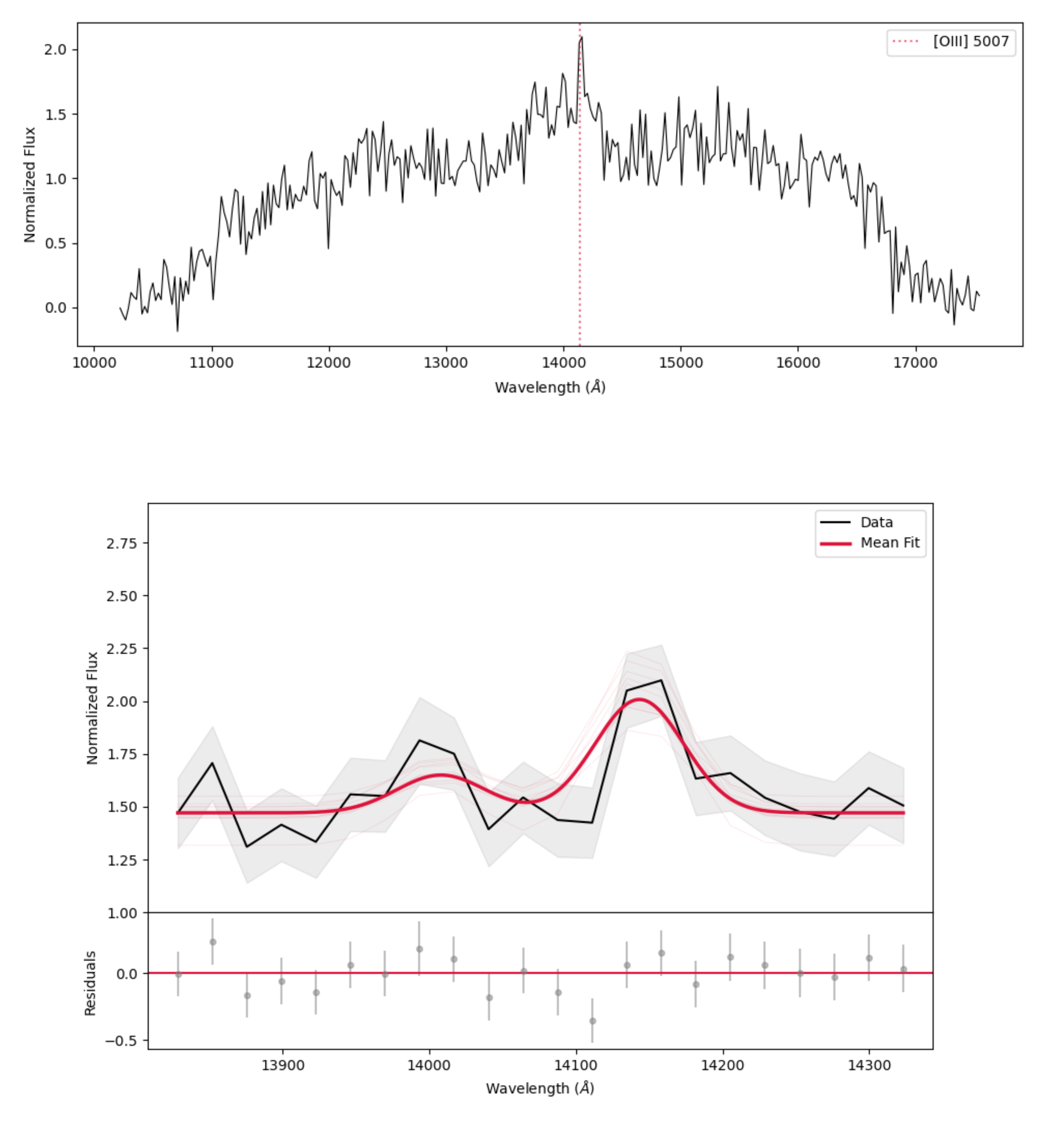}
\caption{Top: HST WFC1/G141 spectrum. The vertical red line represents the location of the redshifted [\ion{O}{3}] line. Bottom: Zoomed-in [\ion{O}{3}] line region. Light red curves represent 100 iterations of the least-squares fit, and the solid red curve represents the mean fit.
\label{fig:general2}}
\end{figure}

\section{data and analysis} \label{sec:style}
\subsection{The Minor Merger}
The system was identified in the 3D-HST survey, a spectroscopic program conducted with the Hubble Space Telescope that combines WFC3/G141 grism spectroscopy with multiwavelength photometry across several extragalactic fields \citep{2012ApJS..200...13B,2014ApJS..214...24S,2016ApJS..225...27M}. Redshifts and physical properties are derived through joint fitting of the grism spectra and spectral energy distributions. The merger is located in the UKIDSS Ultra Deep Survey field (UDS), and both components have spectroscopic redshifts: $z_{\rm prim} = 1.8235$ and $z_{\rm sec} = 1.8246$. We adopt the average value of $z=1.824$ as a working value throughout the paper. The redshift offset $\Delta z = 0.0011$ is well within the typical redshift uncertainties of the 3D-HST grism measurements ($\sigma_z \sim 0.003(1+z)$) \citep{2016ApJS..225...27M}, and is therefore consistent with the two galaxies being physically associated. The stellar masses of the primary and secondary components are log M$_*$ = 11.37 and log M $_*$ = 9.82, which implies a mass ratio of $\sim$35:1. Stellar masses are adopted from the 3D-HST Survey, and are derived via broadband spectral energy distribution fitting, using FAST code \citep{2018ascl.soft03008K} with Bruzual \& Charlot \citep{2003MNRAS.344.1000B} templates and a Chabrier initial mass function \citep{2003PASP..115..763C}, and have typical uncertainties of $\sim$0.1-0.3 dex. Deep imaging from the James Webb Space Telescope, shown in the right panel of Figure \ref{fig:general1}, reveals clear tidal debris in the form of tidal tails emanating from both galaxies, providing strong visual evidence for an ongoing interaction and supporting the merger interpretation.
\subsection{Chandra Observations and HST Spectroscopy}
The minor merger was observed 16 times with the Chandra ACIS-I instrument in 2016 (PI: Hasinger). To ensure optimal spatial resolution and sensitivity, we restricted the dataset to observations in which the source lies within an off-axis angle of $<5\arcmin$ from the aim-point, where the point spread function remains sufficiently compact. The data, comprising six observations with a total exposure time of 300 ksec, were retrieved from the archive, reprocessed, and merged using the CIAO software package with the latest calibration database \citep{2006SPIE.6270E..1VF}. The data is contained in DOI: \dataset[10.25574/cdc.582]{https://doi.org/10.25574/cdc.582}.

We make use of near-infrared slitless spectroscopic data obtained with the Wide Field Camera 3 (WFC3) G141 $grism$ as part of the 3D-HST survey. The G141 grism provides low-resolution spectroscopy ($R \sim 130$) over the wavelength range $1.1$–$1.7,\mu$m. In this work, we directly used the publicly available one-dimensional spectra released by the 3D-HST survey. No additional spectral extraction or reprocessing was performed. The data is available at MAST:\dataset[10.17909/ve5z-d119]{\doi{10.17909/ve5z-d119}}.

\section{Results} \label{sec:floats}
\subsection{X-ray Analysis}
We analyzed the merged X-ray data described in Section \ref{sec:style} to characterize the system's emission. Source counts were extracted from a circular region with a radius of 1\farcs65, which corresponds to 90\% of the encircled energy fraction of the point spread function. The background was estimated from an annular, source-free region surrounding the circular region. Within the source aperture, we detect $121 \pm 11$ net X-ray photons. No significant excess is detected at the position of the primary galaxy, whose counts are consistent with the local background. The left panel of Figure \ref{fig:general1} shows the Chandra image in the 0.2-6 keV band, while the right panel shows X-ray contours superposed onto the JWST image of a minor merger.

Spectral products were generated for individual observations and then combined using the CIAO's \texttt{specextract} and \texttt{combine\_spectra} procedures. Spectral fitting was performed using Sherpa \citep{2001SPIE.4477...76F,2024ApJS..274...43S}. The data were modeled with an absorbed power-law of the form \texttt{phabs*zpowerlw}, where the absorption component was fixed to the galactic column density. The best-fit model yields an observed 0.2-6 keV flux of 3.7$\pm$0.4 $\times$ 10$^{-17}$ erg s$^{-1}$ cm$^{-2}$ and a corresponding luminosity of 9.0$\pm$0.1 $\times$ 10$^{43}$ erg s$^{-1}$ in the rest-frame 2–10 keV band, and a photon index of $\Gamma = 2.0\pm0.2$.
\subsection{HST Spectral Fitting}
We decomposed the WFC3/G141 grism spectra to characterize the [\ion{O}{3}] $\lambda\lambda 4959, 5007$ doublet using a multi-component Gaussian profile superimposed on a local linear continuum. To ensure a robust fit, we kinematically tied the doublet components, enforcing a common velocity dispersion and a fixed theoretical flux ratio of $f_{4959}/f_{5007} = 0.33$. The uncertainties of the parameters were estimated using a Monte Carlo approach, where we performed 100 iterations of the Levenberg-Marquardt least-squares fit on the observed flux perturbed by the local $1\sigma$ error envelope.

The total integrated [\ion{O}{3}] $\lambda 5007$ flux was derived from the best-fit Gaussian amplitude and width, with the signal-to-noise ratio defined by the mean flux relative to the standard deviation of the Monte Carlo distribution. We find the flux of 8.5$\pm$2.2$\times$10$^{-17}$ erg s$^{-1}$ cm$^{-2}$, corresponding to the luminosity of 2.0$\pm$0.5$\times$10$^{42}$ erg s$^{-1}$. The full WFC3/G141 spectra and the fitted [\ion{O}{3}] doublet region are shown in Figure \ref{fig:general2}
\subsection{Bolometric Luminosity} 
\label{subsec:tables}
Estimating the bolometric luminosity ($L_{\rm bol}$) is essential for assessing the total accretion power of the AGN and placing it in the context of black hole growth. Because direct measurements of the full spectral energy distribution are not available, we infer $L_{\rm bol}$ using empirical bolometric corrections ($k_{bol}$) based on both the X-ray and [\ion{O}{3}] emission.

The $L_{\rm bol}$ was derived from the 2-10 X-ray luminosity by applying a standard luminosity-dependent bolometric correction:
\begin{equation}
    \textup{log}(L_{\rm bol}/L_{2-10,keV})=1.54+0.24\mathcal{L}+0.012\mathcal{L}^2-0.0015\mathcal{L}^3
\end{equation}
where $\mathcal{L}$=log $L_{\rm bol}$-12, and $L_{\rm bol}$ expressed in $L_{\odot}$ \citep{2004MNRAS.351..169M}. The resulting bolometric correction is $k_{bol}\approx32$, and bolometric luminosity $L_{\rm bol} = 2.9^{+0.7}_{-0.6} \times 10^{45}$ erg s$^{-1}$, with quoted uncertainties reflecting the measurement uncertainty on $L_{2-10,keV}$ and the intrinsic scatter of the bolometric correction relation ($\pm$0.1 dex).

Since the [\ion{O}{3}] emission originates in the narrow-line region photoionized by the AGN ionizing continuum, it provides an indirect tracer of the bolometric luminosity, complementary to X-ray measurements. The bolometric luminosity is estimated using the commonly adopted relation:
\begin{equation}
    L_{\rm bol} \approx 3500L_{\rm [OIII]}
\end{equation}
,which yields $L_{\rm bol} = 7.0^{+10.4}_{-4.2} \times 10^{45}$ erg s$^{-1}$. The quoted uncertainties reflect the measurement uncertainty on [\ion{O}{3}] and the intrinsic scatter of the bolometric correction relation ($\pm$0.38 dex). Therefore, we conclude that the [\ion{O}{3}]-based and X-ray-based bolometric luminosities are in good agreement.

We also consider luminosity-dependent corrections, which account for changes in the narrow-line region covering fraction with AGN luminosity, given as:
\begin{equation}
    L_{bol}/10^{43}\, \textup{erg\,s}^{-1}=4000(L_{\rm [OIII]}/10^{43}\, \textup{erg\,s}^{-1})^{1.39}
\end{equation}
This relation yields bolometric luminosity of $L_{\rm bol} = 4.3^{+13.5}_{-3.3} \times 10^{45}$ erg s$^{-1}$, in good agreement with previously derived bolometric luminosities. The quoted uncertainties reflect the measurement uncertainty on [\ion{O}{3}] and the intrinsic scatter of the bolometric correction relation ($\pm$0.6 dex) \citep{2012MNRAS.426.2703S}.
\subsection{Black Hole Mass and Accretion Rates}
To place the secondary AGN in the context of black hole–galaxy scaling relations, we estimate the expected black hole mass from the stellar mass of the host galaxy:
\begin{equation}
    \textup{log}(M_{BH}/M_{\odot})=7.45\pm0.08+1.05\pm0.11\textup{log}(M_*/10^{11}M_{\odot})
\end{equation}
This relation was constructed using a sample of galaxies with stellar masses $10^8\leq M_*/M_{\odot}\leq10^{12}$, and is the most reliable available black hole mass estimator since the secondary galaxy, at log M $_*$ = 9.82, is a relatively low-mass galaxy \citep{2015ApJ...813...82R}. The resulting black hole mass is $M_{\rm BH} = 1.6^{+5.7}_{-1.3} \times 10^{6}\ M_{\odot}$, where the uncertainty accounts for the errors on the relation coefficients, the uncertainty on the host galaxy stellar mass ($\pm$0.3 dex), and the intrinsic scatter of the relation ($\pm$0.55 dex), all added in quadrature. The black hole mass scales with Eddington luminosity via the following equation:
\begin{equation}
    L_{Edd}\approx1.26\times10^{38}(M_{BH}/M_{\odot})
\end{equation}
, where Eddington luminosity can be rewritten in terms of bolometric luminosity as L$_{Edd}$=L$_{bol}$/$\lambda_{Edd}$, where $\lambda_{Edd}$ is the Eddington ratio. Substituting the previously derived black hole mass, we find Eddington luminosity of L$_{Edd}$=2.1$^{+7.1}_{-1.6}\times$10$^{44}$ erg s$^{-1}$. Adopting the X-ray-based bolometric luminosity as our primary estimate, as X-ray emission provides the most direct tracer of AGN accretion power and the most conservative estimate, requires an Eddington ratio in the range of $\lambda_{Edd}$=3-70. While such extreme super-Eddington accretion is theoretically possible, it is typically expected to arise from stellar-mass black holes or tidal disruption events \citep{2009MNRAS.400.2070S,2014Natur.514..202B}. Therefore, we conclude that it is physically unlikely that a regular-sized SMBH is powering the secondary AGN.

On the other hand, assuming Eddington-limited accretion ($\lambda_{Edd}$=1), a physically plausible scenario, implies a black hole mass of $M_{\rm BH} = 2.3^{+0.6}_{-0.5} \times 10^{7}\ M_{\odot}$. This mass is an order of magnitude more massive than what is predicted by the M$_*$-M$_{BH}$ relation, and is $\sim$0.35$\%$ of the secondary galaxy stellar mass. For sub-Eddington accretion, the required black hole mass scales inversely with the Eddington ratio. As a result, decreasing $\lambda_{\rm Edd}$ drives the inferred black hole mass to values increasingly overmassive from expectations based on the host galaxy stellar mass. For example, assuming a moderately accreting black hole with $\lambda_{Edd}$=0.01, yields a black hole mass that accounts for $\sim$35$\%$ of the host galaxy stellar mass, a physically unlikely scenario.

This coupling between $M_{\rm BH}$ and $\lambda_{\rm Edd}$ introduces a degeneracy: the observed luminosity can be explained either by a lower-mass, but still overmassive, black hole accreting near or at the Eddington limit or by a more overmassive black hole accreting at lower, but still high, rates. Therefore, both scenarios imply that the system deviates from typical scaling relations and that the minor merger secondary AGN is powered by an overmassive, highly accreting, rapidly growing black hole.

Additional constraints can be obtained from the X-ray spectral slope. The photon index $\Gamma$ scales with Eddington ratio, with increasing value of $\Gamma$ corresponding to higher accretion rates. We compare the previously derived photon index of $\Gamma$=2.0$\pm$0.2 with various $\Gamma$-$\lambda_{Edd}$ relations from the literature, constructed using high luminosity AGN and quasars \citep{2008ApJ...682...81S,2013MNRAS.433.2485B}, and find high accretion rate, $\lambda_{Edd}$=0.15-0.25. For this range of Eddington ratios, we find a black hole mass of $\sim$1.5-3$\%$ of the secondary galaxy stellar mass. In addition, we refit the X-ray spectrum, including the free intrinsic absorption component, to test the robustness of $\Gamma$ to the assumed absorption model. Inclusion of intrinsic absorption slightly shifts the photon index to softer values, $\Gamma$ = 2.3$\pm$0.3. The two values are consistent within uncertainties, and the steeper photon index obtained when accounting for intrinsic absorption reinforces the conclusion that this AGN is accreting at a high Eddington ratio.
\section{Summary}
We reported the discovery of a cosmic noon, $z=1.824$, minor merger with mass ratio 35:1. The smaller, secondary galaxy contains a bright, X-ray point source. We performed an analysis of archival X-ray observations and an HST WFC3/G141 spectrum and found a high bolometric luminosity of 3-7$\times$10$^{45}$ erg s$^{-1}$. Assuming a scenario in which the black hole is accreting at the Eddington limit, the black hole mass is by an order of magnitude more massive than what is predicted by the host galaxy stellar mass-black hole mass scaling relation. Assuming sub-Eddington accretion further increases the necessary black hole mass. Therefore, while the exact accretion rate and black hole mass cannot be determined with the current data, we conclude that the secondary AGN must be powered by an overmassive black hole accreting at a moderate to high rate.

\bibliography{sample701}{}

\begin{thebibliography}{}
\expandafter\ifx\csname natexlab\endcsname\relax\def\natexlab#1{#1}\fi
\providecommand{\url}[1]{\href{#1}{#1}}
\providecommand{\dodoi}[1]{doi:~\href{http://doi.org/#1}{\nolinkurl{#1}}}
\providecommand{\doeprint}[1]{\href{http://ascl.net/#1}{\nolinkurl{http://ascl.net/#1}}}
\providecommand{\doarXiv}[1]{\href{https://arxiv.org/abs/#1}{\nolinkurl{https://arxiv.org/abs/#1}}}

\bibitem[{E. {Ba{\~n}ados} {et~al.}(2018){Ba{\~n}ados}, {Venemans}, {Mazzucchelli}, {Farina}, {Walter}, {Wang}, {Decarli}, {Stern}, {Fan}, {Davies}, {Hennawi}, {Simcoe}, {Turner}, {Rix}, {Yang}, {Kelson}, {Rudie}, \& {Winters}}]{2018Natur.553..473B}
{Ba{\~n}ados}, E., {Venemans}, B.~P., {Mazzucchelli}, C., {et~al.} 2018, \bibinfo{title}{{An 800-million-solar-mass black hole in a significantly neutral Universe at a redshift of 7.5},} \nat, 553, 473, \dodoi{10.1038/nature25180}

\bibitem[{M. {Bachetti} {et~al.}(2014){Bachetti}, {Harrison}, {Walton}, {Grefenstette}, {Chakrabarty}, {F{\"u}rst}, {Barret}, {Beloborodov}, {Boggs}, {Christensen}, {Craig}, {Fabian}, {Hailey}, {Hornschemeier}, {Kaspi}, {Kulkarni}, {Maccarone}, {Miller}, {Rana}, {Stern}, {Tendulkar}, {Tomsick}, {Webb}, \& {Zhang}}]{2014Natur.514..202B}
{Bachetti}, M., {Harrison}, F.~A., {Walton}, D.~J., {et~al.} 2014, \bibinfo{title}{{An ultraluminous X-ray source powered by an accreting neutron star},} \nat, 514, 202, \dodoi{10.1038/nature13791}

\bibitem[{{\'A}. {Bogd{\'a}n} {et~al.}(2024){Bogd{\'a}n}, {Goulding}, {Natarajan}, {Kov{\'a}cs}, {Tremblay}, {Chadayammuri}, {Volonteri}, {Kraft}, {Forman}, {Jones}, {Churazov}, \& {Zhuravleva}}]{2024NatAs...8..126B}
{Bogd{\'a}n}, {\'A}., {Goulding}, A.~D., {Natarajan}, P., {et~al.} 2024, \bibinfo{title}{{Evidence for heavy-seed origin of early supermassive black holes from a z ≍ 10 X-ray quasar},} Nature Astronomy, 8, 126, \dodoi{10.1038/s41550-023-02111-9}

\bibitem[{G.~D. {Bothun} {et~al.}(1989){Bothun}, {Halpern}, {Lonsdale}, {Impey}, \& {Schmitz}}]{1989ApJS...70..271B}
{Bothun}, G.~D., {Halpern}, J.~P., {Lonsdale}, C.~J., {Impey}, C., \& {Schmitz}, M. 1989, \bibinfo{title}{{The Wasilewski Sample of Emission-Line Galaxies: Follow-up CCD Imaging and Spectroscopic and IRAS Observations},} \apjs, 70, 271, \dodoi{10.1086/191341}

\bibitem[{G.~B. {Brammer} {et~al.}(2012){Brammer}, {van Dokkum}, {Franx}, {Fumagalli}, {Patel}, {Rix}, {Skelton}, {Kriek}, {Nelson}, {Schmidt}, {Bezanson}, {da Cunha}, {Erb}, {Fan}, {F{\"o}rster Schreiber}, {Illingworth}, {Labb{\'e}}, {Leja}, {Lundgren}, {Magee}, {Marchesini}, {McCarthy}, {Momcheva}, {Muzzin}, {Quadri}, {Steidel}, {Tal}, {Wake}, {Whitaker}, \& {Williams}}]{2012ApJS..200...13B}
{Brammer}, G.~B., {van Dokkum}, P.~G., {Franx}, M., {et~al.} 2012, \bibinfo{title}{{3D-HST: A Wide-field Grism Spectroscopic Survey with the Hubble Space Telescope},} \apjs, 200, 13, \dodoi{10.1088/0067-0049/200/2/13}

\bibitem[{M. {Brightman} {et~al.}(2013){Brightman}, {Silverman}, {Mainieri}, {Ueda}, {Schramm}, {Matsuoka}, {Nagao}, {Steinhardt}, {Kartaltepe}, {Sanders}, {Treister}, {Shemmer}, {Brandt}, {Brusa}, {Comastri}, {Ho}, {Lanzuisi}, {Lusso}, {Nandra}, {Salvato}, {Zamorani}, {Akiyama}, {Alexander}, {Bongiorno}, {Capak}, {Civano}, {Del Moro}, {Doi}, {Elvis}, {Hasinger}, {Laird}, {Masters}, {Mignoli}, {Ohta}, {Schawinski}, \& {Taniguchi}}]{2013MNRAS.433.2485B}
{Brightman}, M., {Silverman}, J.~D., {Mainieri}, V., {et~al.} 2013, \bibinfo{title}{{A statistical relation between the X-ray spectral index and Eddington ratio of active galactic nuclei in deep surveys},} \mnras, 433, 2485, \dodoi{10.1093/mnras/stt920}

\bibitem[{G. {Bruzual} \& S. {Charlot}(2003){Bruzual} \& {Charlot}}]{2003MNRAS.344.1000B}
{Bruzual}, G., \& {Charlot}, S. 2003, \bibinfo{title}{{Stellar population synthesis at the resolution of 2003},} \mnras, 344, 1000, \dodoi{10.1046/j.1365-8711.2003.06897.x}

\bibitem[{S. {Callegari} {et~al.}(2011){Callegari}, {Kazantzidis}, {Mayer}, {Colpi}, {Bellovary}, {Quinn}, \& {Wadsley}}]{2011ApJ...729...85C}
{Callegari}, S., {Kazantzidis}, S., {Mayer}, L., {et~al.} 2011, \bibinfo{title}{{Growing Massive Black Hole Pairs in Minor Mergers of Disk Galaxies},} \apj, 729, 85, \dodoi{10.1088/0004-637X/729/2/85}

\bibitem[{P.~R. {Capelo} {et~al.}(2015){Capelo}, {Volonteri}, {Dotti}, {Bellovary}, {Mayer}, \& {Governato}}]{2015MNRAS.447.2123C}
{Capelo}, P.~R., {Volonteri}, M., {Dotti}, M., {et~al.} 2015, \bibinfo{title}{{Growth and activity of black holes in galaxy mergers with varying mass ratios},} \mnras, 447, 2123, \dodoi{10.1093/mnras/stu2500}

\bibitem[{G. {Chabrier}(2003){Chabrier}}]{2003PASP..115..763C}
{Chabrier}, G. 2003, \bibinfo{title}{{Galactic Stellar and Substellar Initial Mass Function},} \pasp, 115, 763, \dodoi{10.1086/376392}

\bibitem[{J.~M. {Comerford} {et~al.}(2015){Comerford}, {Pooley}, {Barrows}, {Greene}, {Zakamska}, {Madejski}, \& {Cooper}}]{2015ApJ...806..219C}
{Comerford}, J.~M., {Pooley}, D., {Barrows}, R.~S., {et~al.} 2015, \bibinfo{title}{{Merger-driven Fueling of Active Galactic Nuclei: Six Dual and Offset AGNs Discovered with Chandra and Hubble Space Telescope Observations},} \apj, 806, 219, \dodoi{10.1088/0004-637X/806/2/219}

\bibitem[{S.~L. {Ellison} {et~al.}(2013){Ellison}, {Mendel}, {Patton}, \& {Scudder}}]{2013MNRAS.435.3627E}
{Ellison}, S.~L., {Mendel}, J.~T., {Patton}, D.~R., \& {Scudder}, J.~M. 2013, \bibinfo{title}{{Galaxy pairs in the Sloan Digital Sky Survey - VIII. The observational properties of post-merger galaxies},} \mnras, 435, 3627, \dodoi{10.1093/mnras/stt1562}

\bibitem[{S.~L. {Ellison} {et~al.}(2011){Ellison}, {Patton}, {Mendel}, \& {Scudder}}]{2011MNRAS.418.2043E}
{Ellison}, S.~L., {Patton}, D.~R., {Mendel}, J.~T., \& {Scudder}, J.~M. 2011, \bibinfo{title}{{Galaxy pairs in the Sloan Digital Sky Survey - IV. Interactions trigger active galactic nuclei},} \mnras, 418, 2043, \dodoi{10.1111/j.1365-2966.2011.19624.x}

\bibitem[{P. {Freeman} {et~al.}(2001){Freeman}, {Doe}, \& {Siemiginowska}}]{2001SPIE.4477...76F}
{Freeman}, P., {Doe}, S., \& {Siemiginowska}, A. 2001, \bibinfo{title}{{Sherpa: a mission-independent data analysis application},} in Society of Photo-Optical Instrumentation Engineers (SPIE) Conference Series, Vol. 4477, Astronomical Data Analysis, ed. J.-L. {Starck} \& F.~D. {Murtagh}, 76--87, \dodoi{10.1117/12.447161}

\bibitem[{A. {Fruscione} {et~al.}(2006){Fruscione}, {McDowell}, {Allen}, {Brickhouse}, {Burke}, {Davis}, {Durham}, {Elvis}, {Galle}, {Harris}, {Huenemoerder}, {Houck}, {Ishibashi}, {Karovska}, {Nicastro}, {Noble}, {Nowak}, {Primini}, {Siemiginowska}, {Smith}, \& {Wise}}]{2006SPIE.6270E..1VF}
{Fruscione}, A., {McDowell}, J.~C., {Allen}, G.~E., {et~al.} 2006, \bibinfo{title}{{CIAO: Chandra's data analysis system},} in Society of Photo-Optical Instrumentation Engineers (SPIE) Conference Series, Vol. 6270, Observatory Operations: Strategies, Processes, and Systems, ed. D.~R. {Silva} \& R.~E. {Doxsey}, 62701V, \dodoi{10.1117/12.671760}

\bibitem[{F. {Gao} {et~al.}(2020){Gao}, {Wang}, {Pearson}, {Gordon}, {Holwerda}, {Hopkins}, {Brown}, {Bland-Hawthorn}, \& {Owers}}]{2020A&A...637A..94G}
{Gao}, F., {Wang}, L., {Pearson}, W.~J., {et~al.} 2020, \bibinfo{title}{{Mergers trigger active galactic nuclei out to z {\ensuremath{\sim}} 0.6},} \aap, 637, A94, \dodoi{10.1051/0004-6361/201937178}

\bibitem[{A.~D. {Goulding} {et~al.}(2018){Goulding}, {Greene}, {Bezanson}, {Greco}, {Johnson}, {Leauthaud}, {Matsuoka}, {Medezinski}, \& {Price-Whelan}}]{2018PASJ...70S..37G}
{Goulding}, A.~D., {Greene}, J.~E., {Bezanson}, R., {et~al.} 2018, \bibinfo{title}{{Galaxy interactions trigger rapid black hole growth: An unprecedented view from the Hyper Suprime-Cam survey},} \pasj, 70, S37, \dodoi{10.1093/pasj/psx135}

\bibitem[{M. {Koss} {et~al.}(2012){Koss}, {Mushotzky}, {Treister}, {Veilleux}, {Vasudevan}, \& {Trippe}}]{2012ApJ...746L..22K}
{Koss}, M., {Mushotzky}, R., {Treister}, E., {et~al.} 2012, \bibinfo{title}{{Understanding Dual Active Galactic Nucleus Activation in the nearby Universe},} \apjl, 746, L22, \dodoi{10.1088/2041-8205/746/2/L22}

\bibitem[{M. {Kriek} {et~al.}(2018){Kriek}, {van Dokkum}, {Labb{\'e}}, {Franx}, {Illingworth}, {Marchesini}, {Quadri}, {Aird}, {Coil}, \& {Georgakakis}}]{2018ascl.soft03008K}
{Kriek}, M., {van Dokkum}, P.~G., {Labb{\'e}}, I., {et~al.} 2018, {FAST: Fitting and Assessment of Synthetic Templates},, Astrophysics Source Code Library, record ascl:1803.008 \doeprint{1803.008}

\bibitem[{X. {Liu} {et~al.}(2018){Liu}, {Guo}, {Shen}, {Greene}, \& {Strauss}}]{2018ApJ...862...29L}
{Liu}, X., {Guo}, H., {Shen}, Y., {Greene}, J.~E., \& {Strauss}, M.~A. 2018, \bibinfo{title}{{Hubble Space Telescope Wide Field Camera 3 Identifies an r $_{ p }$ = 1 Kpc Dual Active Galactic Nucleus in the Minor Galaxy Merger SDSS J0924+0510 at z = 0.1495},} \apj, 862, 29, \dodoi{10.3847/1538-4357/aac9cb}

\bibitem[{A. {Marconi} {et~al.}(2004){Marconi}, {Risaliti}, {Gilli}, {Hunt}, {Maiolino}, \& {Salvati}}]{2004MNRAS.351..169M}
{Marconi}, A., {Risaliti}, G., {Gilli}, R., {et~al.} 2004, \bibinfo{title}{{Local supermassive black holes, relics of active galactic nuclei and the X-ray background},} \mnras, 351, 169, \dodoi{10.1111/j.1365-2966.2004.07765.x}

\bibitem[{M. {Mi{\'c}i{\'c}} {et~al.}(2024{\natexlab{a}}){Mi{\'c}i{\'c}}, {Irwin}, {Nair}, {Wells}, {Holmes}, \& {Eames}}]{2024ApJ...968L..21M}
{Mi{\'c}i{\'c}}, M., {Irwin}, J.~A., {Nair}, P., {et~al.} 2024{\natexlab{a}}, \bibinfo{title}{{Low-mass Galaxy Interactions Trigger Black Hole Activity},} \apjl, 968, L21, \dodoi{10.3847/2041-8213/ad5345}

\bibitem[{M. {Mi{\'c}i{\'c}} {et~al.}(2024{\natexlab{b}}){Mi{\'c}i{\'c}}, {Wells}, {Holmes}, \& {Irwin}}]{2024OJAp....7E...3M}
{Mi{\'c}i{\'c}}, M., {Wells}, B.~N., {Holmes}, O.~J., \& {Irwin}, J.~A. 2024{\natexlab{b}}, \bibinfo{title}{{SDSS J125417.98+274004.6: An X-ray Detected Minor Merger Dual AGN},} The Open Journal of Astrophysics, 7, 3, \dodoi{10.21105/astro.2310.09945}

\bibitem[{I.~G. {Momcheva} {et~al.}(2016){Momcheva}, {Brammer}, {van Dokkum}, {Skelton}, {Whitaker}, {Nelson}, {Fumagalli}, {Maseda}, {Leja}, {Franx}, {Rix}, {Bezanson}, {Da Cunha}, {Dickey}, {F{\"o}rster Schreiber}, {Illingworth}, {Kriek}, {Labb{\'e}}, {Ulf Lange}, {Lundgren}, {Magee}, {Marchesini}, {Oesch}, {Pacifici}, {Patel}, {Price}, {Tal}, {Wake}, {van der Wel}, \& {Wuyts}}]{2016ApJS..225...27M}
{Momcheva}, I.~G., {Brammer}, G.~B., {van Dokkum}, P.~G., {et~al.} 2016, \bibinfo{title}{{The 3D-HST Survey: Hubble Space Telescope WFC3/G141 Grism Spectra, Redshifts, and Emission Line Measurements for \raisebox{-0.5ex}\textasciitilde 100,000 Galaxies},} \apjs, 225, 27, \dodoi{10.3847/0067-0049/225/2/27}

\bibitem[{E.~C. {Moran} {et~al.}(1992){Moran}, {Halpern}, {Bothun}, \& {Becker}}]{1992AJ....104..990M}
{Moran}, E.~C., {Halpern}, J.~P., {Bothun}, G.~D., \& {Becker}, R.~H. 1992, \bibinfo{title}{{WAS 49: Mirror for a Hidden Seyfert 1 Nucleus},} \aj, 104, 990, \dodoi{10.1086/116292}

\bibitem[{A.~E. {Reines} \& M. {Volonteri}(2015){Reines} \& {Volonteri}}]{2015ApJ...813...82R}
{Reines}, A.~E., \& {Volonteri}, M. 2015, \bibinfo{title}{{Relations between Central Black Hole Mass and Total Galaxy Stellar Mass in the Local Universe},} \apj, 813, 82, \dodoi{10.1088/0004-637X/813/2/82}

\bibitem[{S. {Satyapal} {et~al.}(2014){Satyapal}, {Ellison}, {McAlpine}, {Hickox}, {Patton}, \& {Mendel}}]{2014MNRAS.441.1297S}
{Satyapal}, S., {Ellison}, S.~L., {McAlpine}, W., {et~al.} 2014, \bibinfo{title}{{Galaxy pairs in the Sloan Digital Sky Survey - IX. Merger-induced AGN activity as traced by the Wide-field Infrared Survey Explorer},} \mnras, 441, 1297, \dodoi{10.1093/mnras/stu650}

\bibitem[{N.~J. {Secrest} {et~al.}(2017){Secrest}, {Schmitt}, {Blecha}, {Rothberg}, \& {Fischer}}]{2017ApJ...836..183S}
{Secrest}, N.~J., {Schmitt}, H.~R., {Blecha}, L., {Rothberg}, B., \& {Fischer}, J. 2017, \bibinfo{title}{{Was 49b: An Overmassive AGN in a Merging Dwarf Galaxy?},} \apj, 836, 183, \dodoi{10.3847/1538-4357/836/2/183}

\bibitem[{O. {Shemmer} {et~al.}(2008){Shemmer}, {Brandt}, {Netzer}, {Maiolino}, \& {Kaspi}}]{2008ApJ...682...81S}
{Shemmer}, O., {Brandt}, W.~N., {Netzer}, H., {Maiolino}, R., \& {Kaspi}, S. 2008, \bibinfo{title}{{The Hard X-Ray Spectrum as a Probe for Black Hole Growth in Radio-Quiet Active Galactic Nuclei},} \apj, 682, 81, \dodoi{10.1086/588776}

\bibitem[{A. {Siemiginowska} {et~al.}(2024){Siemiginowska}, {Burke}, {G{\"u}nther}, {Lee}, {McLaughlin}, {Principe}, {Cheer}, {Fruscione}, {Laurino}, {McDowell}, \& {Terrell}}]{2024ApJS..274...43S}
{Siemiginowska}, A., {Burke}, D., {G{\"u}nther}, H.~M., {et~al.} 2024, \bibinfo{title}{{Sherpa: An Open-source Python Fitting Package},} \apjs, 274, 43, \dodoi{10.3847/1538-4365/ad7bab}

\bibitem[{R.~E. {Skelton} {et~al.}(2014){Skelton}, {Whitaker}, {Momcheva}, {Brammer}, {van Dokkum}, {Labb{\'e}}, {Franx}, {van der Wel}, {Bezanson}, {Da Cunha}, {Fumagalli}, {F{\"o}rster Schreiber}, {Kriek}, {Leja}, {Lundgren}, {Magee}, {Marchesini}, {Maseda}, {Nelson}, {Oesch}, {Pacifici}, {Patel}, {Price}, {Rix}, {Tal}, {Wake}, \& {Wuyts}}]{2014ApJS..214...24S}
{Skelton}, R.~E., {Whitaker}, K.~E., {Momcheva}, I.~G., {et~al.} 2014, \bibinfo{title}{{3D-HST WFC3-selected Photometric Catalogs in the Five CANDELS/3D-HST Fields: Photometry, Photometric Redshifts, and Stellar Masses},} \apjs, 214, 24, \dodoi{10.1088/0067-0049/214/2/24}

\bibitem[{J. {Stern} \& A. {Laor}(2012){Stern} \& {Laor}}]{2012MNRAS.426.2703S}
{Stern}, J., \& {Laor}, A. 2012, \bibinfo{title}{{Type 1 AGN at low z - II. The relative strength of narrow lines and the nature of intermediate type AGN},} \mnras, 426, 2703, \dodoi{10.1111/j.1365-2966.2012.21772.x}

\bibitem[{L.~E. {Strubbe} \& E. {Quataert}(2009){Strubbe} \& {Quataert}}]{2009MNRAS.400.2070S}
{Strubbe}, L.~E., \& {Quataert}, E. 2009, \bibinfo{title}{{Optical flares from the tidal disruption of stars by massive black holes},} \mnras, 400, 2070, \dodoi{10.1111/j.1365-2966.2009.15599.x}

\bibitem[{S. {Van Wassenhove} {et~al.}(2012){Van Wassenhove}, {Volonteri}, {Mayer}, {Dotti}, {Bellovary}, \& {Callegari}}]{2012ApJ...748L...7V}
{Van Wassenhove}, S., {Volonteri}, M., {Mayer}, L., {et~al.} 2012, \bibinfo{title}{{Observability of Dual Active Galactic Nuclei in Merging Galaxies},} \apjl, 748, L7, \dodoi{10.1088/2041-8205/748/1/L7}

\bibitem[{F. {Wang} {et~al.}(2021){Wang}, {Yang}, {Fan}, {Hennawi}, {Barth}, {Banados}, {Bian}, {Boutsia}, {Connor}, {Davies}, {Decarli}, {Eilers}, {Farina}, {Green}, {Jiang}, {Li}, {Mazzucchelli}, {Nanni}, {Schindler}, {Venemans}, {Walter}, {Wu}, \& {Yue}}]{2021ApJ...907L...1W}
{Wang}, F., {Yang}, J., {Fan}, X., {et~al.} 2021, \bibinfo{title}{{A Luminous Quasar at Redshift 7.642},} \apjl, 907, L1, \dodoi{10.3847/2041-8213/abd8c6}

\bibitem[{A.~J. {Wasilewski}(1983){Wasilewski}}]{1983ApJ...272...68W}
{Wasilewski}, A.~J. 1983, \bibinfo{title}{{The space density and spectroscopic properties of a new sample of emission-line galaxies.},} \apj, 272, 68, \dodoi{10.1086/161262}

\bibitem[{J. {Yang} {et~al.}(2020){Yang}, {Wang}, {Fan}, {Hennawi}, {Davies}, {Yue}, {Banados}, {Wu}, {Venemans}, {Barth}, {Bian}, {Boutsia}, {Decarli}, {Farina}, {Green}, {Jiang}, {Li}, {Mazzucchelli}, \& {Walter}}]{2020ApJ...897L..14Y}
{Yang}, J., {Wang}, F., {Fan}, X., {et~al.} 2020, \bibinfo{title}{{P{\={o}}niu{\={a}}'ena: A Luminous z = 7.5 Quasar Hosting a 1.5 Billion Solar Mass Black Hole},} \apjl, 897, L14, \dodoi{10.3847/2041-8213/ab9c26}

\end{thebibliography}
\bibliographystyle{aasjournalv7}



\end{document}